\documentclass[aps,twocolumn,showpacs]{revtex4}
\usepackage{graphics}
\begin{document}
\tighten
\bibliographystyle{apsrev}
\def\half{{1\over 2}}
\def \D {\mbox{D}}
\def\curl {\mbox{curl}\,}
\def \ep {\varepsilon}
\def \lleq {\lower0.9ex\hbox{ $\buildrel < \over \sim$} ~}
\def \ggeq {\lower0.9ex\hbox{ $\buildrel > \over \sim$} ~}
\def\beq{\begin{equation}}
\def\eeq{\end{equation}}
\def\ber{\begin{eqnarray}}
\def\eer{\end{eqnarray}}
\def \apl {ApJ, }
\def \aps {ApJS, }
\def \pd {Phys. Rev. D, }
\def \prl {Phys. Rev. Lett., }
\def \pl {Phys. Lett., }
\def \np {Nucl. Phys., }
\def \l {\Lambda}

%%\textheight 21.0cm
%%\textwidth 16cm
%%\leftmargin -4cm
%%\topmargin 0cm
%%\topmargin -2cm
%\begin{document}
%\draft
\title{Steep Inflation followed by Born-Infeld Reheating}

\author{M.Sami}
\altaffiliation[On leave from:]{ Department of Physics, Jamia Millia, New Delhi-110025}
\email{sami@iucaa.ernet.in}
\author{N. Dadhich} 
\email{nkd@iucaa.ernet.in}
\affiliation{IUCAA, Post Bag 4, Ganeshkhind,\\
 Pune 411 007, India.}
\author{Tetsuya Shiromizu}
\email{shiromizu@th.phys.titech.ac.jp}
\affiliation{Department of Physics, Tokyo Institute of Technology, Tokyo 152-8551, Japan,\\
Advanced Research Institute for Science and Technology, Waseda University, Tokyo 169-8555, Japan.}

%\author{M. Sami$^{\dagger ~ a}$ \altaffiliation[On leave from:]{ Department of Physics, Jamia Millia, New Delhi-110025}  , N. Dadhich$^{\dagger ~ b}$, Tetsuya Shiromizu$^{\ddagger ~ c}$\\
%\small \it $^{\dagger}$ Inter-University Centre for Astronomy and Astrophysics,
%Post Bag 4, Ganeshkhind, Pune-411 007, India. \\
%$^{\ddagger}$ Department of Physics, Tokyo Institute of Technology, Tokyo 152-8551, Japan. \\
%${\ddagger}$ Advanced Research Institute for Science and Technology, Waseda University, Tokyo 169-8555, Japan.}


\begin{abstract}
We discuss a model in which high energy brane corrections allow a single scalar field to describe inflation at early epochs and
quintessence at late times. The reheating mechanism in the model originates from Born-Infeld matter whose energy density 
mimics cosmological constant at very early times and manifests itself as radiation subsequently. For
most of the inflationary evolution the Born-Infeld matter remains subdominant to the  
the scalar field. Shortly before the end of inflation driven by the scalar field, the energy density of Born-Infeld matter starts 
scaling as radiation and drops by several orders of magnitudes at the epoch inflation ends.
The problem of over production
of gravity wave background in scenarios based upon reheating through gravitational particle production is successfully resolved 
by suitably fixing the initial value of radiation energy density at the end of inflation.
No additional fine tuning of the parameters is required for a viable evolution.
\end{abstract}

\maketitle

\section{Introduction}
Recently, there have been attempts to build models with a single scalar field playing the dual role of inflation at early epochs and
quintessence at late times\cite{vilenkin,lid,liddle,lidsey1,tarun,majumdar,lidsey2,nunes}. Apparently, there are two obstacles in carrying out the program of unification of inflation with 
quintessence. First, the conventional reheating mechanism does not work in these models and an alternative reheating method
has to be employed. Secondly, the scalar field after successfully playing the inflaton role should go into hiding so as not to
disturb the thermal history in the standard model. It should emerge out from the shadow only at late times to dominate the energy
density of the universe and become quintessence. As for the reheating, it can be achieved through a quantum mechanical process of
particle production in time varying gravitational field at the end of inflation\cite{ford,spokoiny,liddle}. However, then the inflaton energy density should 
red-shift faster than that of the produced particles so that radiation domination could commence. And this requires a steep scalar
field potential, which of course can not support inflation in the standard FRW cosmology. This is precisely where the brane assisted
inflation comes to the rescue.\par 
The presence of the quadratic density term (high energy corrections) in the Friedman equation on the brane changes the expansion dynamics in the early universe: at early times, the brane world Hubble parameter is much larger than its usual Friedman 
value\cite{randall,
shiromizu,braneindustry}. Consequently, the field experiences greater damping and rolls down its potential slower than it would during the conventional inflation.
Thus, inflation in the brane world scenario can successfully occur for very steep potentials 
(see Ref\cite{gman} for a different approach to steep inflation).\par 
The second problem can be resolved by 
employing the ``Tracker Fields". For instance, scalar field models with exponential potential have an attractive property that the
energy density in the field tracks the background (radiation and/or matter) energy density remaining sub-dominant so as to respect
the nucleo-synthesis constraint. However at late times, the potential should behave otherwise, allowing the scalar field to play the
dominant role in the evolution dynamics of the universe. One, possibility is that during late times the potential changes to power law
, resulting in rapid oscillations of the field near origin. In case of a particular power law, the average equation of state may turn
negative making the field energy density scale slower (than the background energy density) and dominant at late times to account for
the observed acceleration of the universe. The {\it cosine hyperbolic} potentials has both these features. Another possibility is provided
by a class of inverse power law potentials which at late times become shallow enough to support the conventional FRW inflation allowing the 
inflaton to become a `quintessence'. It is remarkable that extra dimensional effects allow 
a single scalar field to play the dual role of inflaton as well as the dark energy.\par 
Unfortunately these models using the gravitational particle production as a reheating mechanism are faced with a serious problem associated with the relic gravity background. In fact, a generic feature of inflationary models with
steep potentials is that the post inflationary epoch is characterized by a prolonged kinetic regime which lasts until the energy
density of radiation (created quantum mechanically during inflation) becomes equal to the scalar field energy density. A long duration of kinetic
regime results in a gravity wave spectral density which increases with wave number for wave lengths shorter than the comoving Hubble
 radius at the commencement of the radiative regime and leads to violation of the cosmological nucleosynthesis constraints\cite{tarun}.\par
An interesting proposal which circumvents this difficulty has recently been suggested by Liddle and Lopez\cite{curvaton}, see also 
Refs\cite{curvaton1,curvatonJ,curvaton2,curvaton3} on the similar theme. The authors have employed
a new method of reheating via curvaton to address the problems associated with gravitational particle
production mechanism. The curvaton model as shown in Ref\cite{curvaton} can in principal resolve the difficulties related to excessive amplitude of
short-scale gravitational waves. Although this model is interesting, it operates through a very complex network of constraints dictated by the fine tuning
of parameters of the model.\par 
A very interesting proposal has recently been made in Ref\cite{bi1}. It is advocated that the Born-Infeld action\cite{bi2,bi3,bi3panda,bi4} should 
be used in the context of the brane world cosmology (see Ref\cite{salim} on the related theme). It is remarkable that the energy density of the Born-Infeld matter mimics the cosmological constant like behavior at very early times and scales as radiation later. 
As the the Born-Infeld action considered in \cite{bi1} is composed of the non-linear vector
fields, the density perturbations are absent in the model. \par 
In this paper we examine an attractive alternative reheating mechanism based upon Born-Infeld action. 
%to the curvaton hypothesis.
We propose a toy model in which apart from the Born-Infeld matter there is a  scalar field with steep potential that drives inflation
on the brane and the field survives to date to become quintessence. The Born-Infeld matter remains subdominant during inflation and does not interfere with the inflationary dynamics of the scalar field. At the end of inflation, the energy density of the Born-Infeld matter behaves
like radiation. The radiation energy density at the end of inflation can be conveniently chosen so as to respect the nucleosynthesis
constraint. Since the Born-Infeld action does not include any new parameter, no extra fine tuning of the parameters is required in this model. \par
 The rest of this paper is organized as follows. In the section II, we briefly review the Born-Infeld 
brane world proposed in Ref \cite{bi1}. In the section III, we would suggest an inflationary scenario 
with the Born-Infeld reheating and discuss the evolution of the universe. Finally we give the summary 
and discussion.

\section{Born-Infeld Brane Worlds}
The D-branes are fundamental objects in string theory. The end points of the open string to which the gauge fields are attached are
constrained to lie on the branes. As the string theory contains gravity, the D-branes are the dynamical objects. The effective 
D-brane action is given by the Born-Infeld action
\begin{equation}
S_{BI}=-\lambda_b \int{d^4x \sqrt{-{\rm det}\left(g_{\mu \nu}+F_{\mu \nu}\right)}}
\label{bi}
\end{equation}
where $F_{\mu \nu}$ is the elecromagnetic field tensor (Non-Abelian gauge fields could also be included in the action) and
$\lambda_b$ is the brane tension. The Born-Infeld action, in general, also includes Fermi fields which have been dropped for simplicity. In the brane 
world scenario {\it a la} Randall-Sundrum one adopts the Nambu-Goto action instead of the Born-Infeld action. Shiromizu et al have suggested that in the true spirit of the string theory, the total action in the brane world cosmology be composed of the bulk and D-brane 
actions
\begin{equation}
S=S_{bulk}+S_{BI},
\end{equation}
 where $S_{bulk}$ is the five dimensional Einstein-Hilbert action with the negative 
cosmological constant.
The stress tensor appearing on the right hand side (RHS) of the Einstein equations on the brane will now be sourced by the Born-Infeld
action. The modified Friedman equation on a spatially flat FRW brane acquires the form
\begin{equation}
H^2={1 \over 3M_p^2} \rho_{BI} \left(1+{\rho_{BI} \over 2\lambda_b} \right)
\label{friedman}
\end{equation}
with $\rho_{BI}$ given by
\begin{equation}
\rho_{BI}=\epsilon+{\epsilon^2 \over {6\lambda_b}}
\end{equation}
where $E^2=B^2=\epsilon$
\footnote{We retain here the terms of the order of $\epsilon^2$ only which is analogous to the 
treatment of the higher order derivative corrections. Although
there should be infinite number of the correction terms, we neglect
them to make the analysis tractable which facilitates the study of entire  history
of our universe. The $\epsilon^2$ term corresponds to the $R^2$ term in the higher derivative theory and 
similarly the higher powers. Our truncation of the series at $\epsilon^2$ is thus similar to the truncation at 
$R^2$ in the higher order theory. }
The tension $\lambda_b$ is tuned so that the 
net cosmological constant on the brane vanishes. 
We have dropped the `dark radiation' term in the equation
 (\ref{friedman}) as it rapidly disappear once inflation
sets in. Spatial averaging is assumed while computing $\rho_{BI}$ 
and $P_{BI}$ from the stress-tensor 
corresponding to action (\ref{bi}). The scaling of energy density of the 
Born-Infeld matter, as usual, can be established from the conservation equation
\begin{equation}
\dot{\rho}_{BI}+3H(\rho_{BI}+P_{BI})=0
\end{equation}
where
\begin{equation}
P_{BI}={\epsilon \over 3}-{\epsilon^2 \over {6\lambda_b}}
\label{pressureeq}
\end{equation}
Interestingly, the pressure due to the Born-Infeld matter becomes negative in the high energy regime allowing the accelerated expansion
at early times without the introduction of a scalar field. As shown in \cite{bi1} , the energy density $\rho_{BI}$ scales as radiation when
$\epsilon << 6\lambda_b$. For $\epsilon> 6\lambda_b$, the Born-Infeld matter energy density starts scaling slowly (logarithmically)
with the scale factor to mimic the cosmological constant like behavior. The point is that the Born-Infeld 
matter is subdominant during  the inflationary stage. It comes to play the important role after the end of inflation
when it behaves like radiation and hence serves as an alternative to reheating mechanism.

\section{The Hybrid Brane Worlds}
The brane world cosmology based upon the Born-Infeld action looks promising as it is perfectly tuned with the D-brane ideology. But
since the Born-Infeld action is composed of the non-linear elecromagnetic field, the D-brane cosmology proposed in 
Ref\cite{bi1} can not accommodate 
density perturbations at least in its present formulation. One could include a scalar field in the Born-Infeld action, say, a tachyon condensate to correct the situation. However, such a scenario faces the difficulties associated with reheating and formation of
acoustics/kinks\cite{tytgat}. We shall therefore not follow this track. We shall assume that the scalar field driving the inflation (quintessence) on the 
brane is described by the usual four dimensional action for the scalar fields. 
The total action is given by 
\begin{equation}
S=S_{bulk}+S_{BI}+S_{\rm 4d-scalar}
\end{equation}
where
\begin{equation}
S_{\rm 4d-scalar}=\int{\left(-{1 \over 2}g^{\mu \nu} \partial_{\mu} \phi \partial_{\nu} \phi-V(\phi)\right) \sqrt{-g}d^4x}
\label{NG}
\end{equation}
The energy momentum tensor for the field $\phi$ which arises from the action (\ref{NG}) is given by
\begin{equation}
T_{\mu \nu}=\partial_{\mu} \phi \partial_{\nu} \phi-g_{\mu \nu} \left[{1 \over 2}g^{\mu \nu} \partial_{\mu} \phi \partial_{\nu} \phi+V(\phi)\right]
\end{equation}
In the homogeneous and isotropic universe, the field energy 
density $\rho_{\phi}$ and pressure $p_{\phi}$ obtained from $T_{\mu \nu}$ are
\begin{equation}
\rho_{\phi}={ \dot{\phi}^2 \over 2}+V(\phi),~~~~p_{\phi}={ \dot{\phi}^2 \over 2}-V(\phi)
\end{equation}
The scalar field propagating on the brane modifies the Friedman equation to
\begin{equation}
H^2={1 \over 3M_p^2} \rho_{tot} \left(1+{\rho_{tot} \over 2\lambda_b} \right)
\label{mfriedman}
\end{equation}
where $\rho_{tot}$ is given by
\begin{equation}
\rho_{tot}=\rho_{\phi}+\rho_{BI}
\end{equation}
The choice of the scalar field potential $V(\phi)$ is dictated by the requirements discussed in the introduction\cite{sahni}

\begin{equation}
V(\phi)=V_0\left[\cosh (\tilde{\alpha}\phi/M_p)-1\right]^p,~~~~p~>0
\label{cosine}
\end{equation}
which has asymptotic forms
\begin{equation}
V(\phi)={V_0 \over 2^p} e^{{\alpha}\phi/M_p},~~~\tilde{\alpha}\phi/M_p~>>1,~~\phi~>0
\label{exppot}
\end{equation}
\begin{equation}
V(\phi)={V_0 \over 2^p}\left(\tilde{\alpha} \phi \over M_p \right)^{2p}~~~~~~|\tilde{\alpha} \phi/M_p|~<<1
\end{equation}
where ${\alpha}=p\tilde{\alpha}$. For $\alpha > \sqrt{2}$, the exponential potential (\ref{exppot}) becomes steep to sustain inflation
in the standard cosmology. However, the increased damping due to the quadratic term in the Friedman equation leads to slow roll inflation at high energies ($V/\lambda_b~>>1$). The number of e-foldings $\cal{N}$, in this model, can be unambiguously fixed to 
${\cal N} \simeq 70$\cite{tarun}.
The COBE normalized value of density perturbations allows to determine the value of the potential at the end of inflation $V_{end}$ and
the brane tension $\lambda_b$\cite{tarun}.
\begin{equation}
V_{end}^{1/4}={{3.26 \times 10^{-4}} \over \alpha} M_p
\end{equation}
\begin{equation}
\lambda_b={{5.88\times10^{-15}M_p^4} \over \alpha^6}
\label{lambda}      
\end{equation}

$V_{end}$ and $\lambda_b$ are related as $V_{end}=2\alpha^2 \lambda_b$. In the scenario based upon reheating via quantum mechanical
particle production during inflation, the radiation density is very small, typically one part in $10^{16}$ and the ratio of the field
energy density to that of radiation has no free parameter to tune. This leads to long kinetic regime which results in an unacceptably
large gravity background\cite{tarun}. The Born-Infeld matter which behaves like radiation (at the end of inflation) has no such problem and can be used for reheating
without conflicting with the nucleosynthesis constraint. Indeed, at the end of inflation $\rho_{BI}$ can be chosen such that $\rho_{BI}^{end}<<6\lambda_b$. We shall confirm that such an initial condition for $\rho_{BI}$ is consistent with the nucleo-synthesis constraint. In that
case the Born-Infeld matter energy density would scale like radiation at the end of inflation. At this epoch the scale factor will be initialized at
$a_{end}=1$. The energy density $\rho_{BI}$ would continue scaling as $1/a^4$ below $a=a_{end}$. The scaling would slow down as $\rho_{BI}$
reaches $6\lambda_b$ which is much smaller than $V_{end}$ for generic steep potentials, say  for $\alpha\ge 5$. Hence $\rho_{BI}$ remains
subdominant to scalar field energy density $\rho_{\phi}$ for the entire inflationary evolution. The Born-Infeld matter comes to play the
important role only at the end of inflation which is in a sense similar to curvaton. But unlike curvaton, it does not
contain any new parameter.\par
After the brane inflation with exponential potential has ended, the brane effects persist for some time and the transition to
kinetic regime is not instantaneous. When the kinetic regime finally commences the temperature of the universe has dropped to\cite{tarun}
\begin{equation}
T_{kin}=T_{end}\left(a_{end} \over a_{kin} \right) = T_{end} F_1(\alpha)
\label{Tkin}
\end{equation}
where $F_1(\alpha)=\left(c+{d\over \alpha^2} \right)$, $c \simeq 0.142$,~ $d \simeq -1.057$ and $T_{end}=\left(\rho_{BI}^{end} \right)^{1/4}$ which will be determined from the nucleo-synthesis constraint.
The equality between scalar field matter and radiation (Born-Infeld matter) takes place at the temperature
\begin{equation}
T_{eq}= T_{end} {F_2(\alpha) \over {\left(\rho_{\phi}/\rho_{BI}\right)_{end}^{1/2}}} 
\label{Teq}
\end{equation}
with $ F_2(\alpha)=\left(e+{f \over \alpha^2} \right)$, $e\simeq 0.0265$,~$f\simeq -0.176$. The 
fitting formulas (\ref{Tkin}) and (\ref{Teq}) are obtained by numerical integration of equations of motion. We once again emphasize
that Born-Infeld matter/radiation does not affect the background spacetime during inflationary stage. The fitting formulas exhibit the post inflationary evolution of $\rho_{\phi}$ (up to the commencement of radiative regime) which is essentially same as discussed in \cite{tarun}.\par

\begin{figure}
\resizebox{3.0in}{!}{\includegraphics{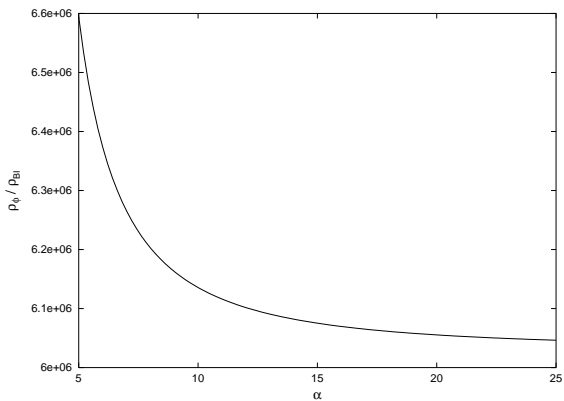}}
\caption{The ratio of field energy density to the radiation density due to Born-Infeld matter at the end of inflation 
is plotted against $\alpha$ ( for the model
of Eq (\ref{exppot})). The ratio is of the order of $10^{7}$ and saturates nearly at $6 \times 10^6$ for large values of $\alpha$.
}
\label{biratio}
\end{figure}

The quantum mechanical production of gravity waves is one of the important aspects of inflationary scenarios\cite{g1,g2,g3,g4,g5,g6,g7,g8}. The relic gravity waves
leave important imprints on the microwave background whose stochastic signature is a challenge to observation. The ratio
of energy density in gravity waves to the radiation energy density at the commencement of radiative regime is given by
\begin{equation}
\left({\rho_g \over \rho_{BI}} \right)_{eq}={64 \over {3 \pi}} h_{GW}^2 \left({T_{kin} \over T_{eq}} \right)^2
\label{graden}
\end{equation}
where $h_{GW}$ is the dimensionless amplitude of gravity waves. From COBE normalization, for ${\cal N} \simeq 70$
\begin{equation}
h_{GW}^2 \simeq 1.7\times 10^{-10}
\end{equation}
Using equations (\ref{Tkin}), (\ref{Teq}) and (\ref{graden}) we obtain the ratio of scalar field energy density to radiation energy
 density (sourced by Born-Infeld matter)  at the end of inflation 
\begin{equation}
{\left(\rho_{\phi} \over \rho_{BI}\right)_{end}}=
{3 \pi \over  64 }\left({1 \over {h_{GW}^2\left(F_1(\alpha)/F_2(\alpha)\right)^2}}\right)\left({\rho_g \over \rho_{BI}} \right)_{eq}
\label{endratio}
\end{equation}
For the nucleo-synthesis constraint to be respected, the ratio of energy density in gravity waves to radiation energy density at equality should obey
$(\rho_g/\rho_{BI})_{eq} \le 0.2$ and this fixes the ratio of the field energy density to the radiation density at the end of inflation.
 It follows from equation (\ref{endratio}) that for generic steep exponential potential ($\alpha\ge5$)
the ratio $\left(\rho_{\phi}/\rho_{BI}\right)_{end} \simeq 10^7$. Since $\rho_{\phi}^{end}\simeq V_{end}=2\alpha^2\lambda_b$, we conclude that $\rho_{BI}^{end} \ll 6\lambda_b$ for a reasonable value of $\alpha \ge 5$. Thus the Born-Infeld
matter behaves like radiation after the end of inflation. A comment on the lower bond of $\alpha$ is,however, in order. As
mentioned in the introduction, 'tracking' is an important component of the post-inflationary behavior of field
energy density $\rho_{\phi}$ in the model under consideration. During the 'tracking regime' $(\rho_{\phi}/\rho_{BI})$
is held fixed
\begin{equation}
{\rho_{\phi} \over {\rho_{BI}+\rho_{\phi}}}={{3(1+w_{BI})} \over \alpha^2} \le 0.2
\label{nuceq}
\end{equation}
The inequality (\ref{nuceq}) reflects the nucleo-synthesis constraint which requires $\alpha$ to be large $\alpha \ge 5$
We once again emphasize that in the gravitational particle production scenario, this ratio is a fixed number (independent of $\alpha$) and can not be
calibrated. The ratio is typically of the order of $10^{16}$ allowing the kinetic regime to be excessively long. Since the gravity wave spectrum is acutely sensitive to the post inflationary equation
of state, a long duration of kinetic regime results in a gravity wave spectral density which increases
with wave number for wavelengths shorter than the comoving horizon scale at the commencement of the
radiative regime. In models based upon reheating via inflationary particle production, the 'blue tilted'
energy density of gravity waves can exceed the radiation density violating the nucleo-synthesis
constraint. The duration of post-inflationary kinetic regime crucially depends upon the ratio of field
energy density to the radiation density at the end of inflation. Interestingly, this ratio is a free parameter
in the scenario based upon reheating with Born-Infeld matter
and can be conveniently set at the end of inflation. Equation (\ref{endratio}) precisely describes the choice of initial value of the ratio
of energy densities $\rho_{\phi}/\rho_{BI}$ such that the evolution would respect the nucleo-synthesis constraint. In the Born-Infeld reheating 
scenario, there is no other free parameter. The only tuning required for a viable evolution is related to the choice of initial condition of the ratio $\rho_{\phi}/\rho_{BI}$. In figure \ref{biratio}, we have plotted this ratio
as a function of $\alpha$ for steep exponential potential (\ref{exppot}). For $\alpha=5$,~$\rho_{\phi}/\rho_{BI} \simeq 6.6 \times 10^{6}$ and the ratio saturates nearly at $6\times 10^6$ for large values of $\alpha$. The radiation density and temperature due to Born-Infeld action can
now be estimated to yield (for $\alpha=5$)
\begin{equation}
\rho_{BI}^{end} \simeq 2.77\times 10^{-24} M_p^4
\label{birad}
\end{equation}
\begin{equation}
T_{end} \simeq  3.14 \times 10^{12} GeV
\end{equation}
\begin{equation}
T_{eq} \simeq 2.38 \times 10^7 GeV
\end{equation}
It should be noted that $T_{eq}$ is larger by several orders of magnitudes than its counter part in the model
based upon reheating via gravitational particle production \cite{tarun}. From Equations (\ref{lambda}) and (\ref{birad}), it is evident that $\rho_{BI} << 6\lambda_b$ at the end of inflation for generic steep potential (\ref{exppot}). Thus the Born-Infeld matter behaves like a perfect radiation after the inflation has ended. During inflation,
as emphasized earlier, the Born-Infeld matter remains subdominant to the scalar field.\par

\begin{figure}
\resizebox{3.0in}{!}{\includegraphics{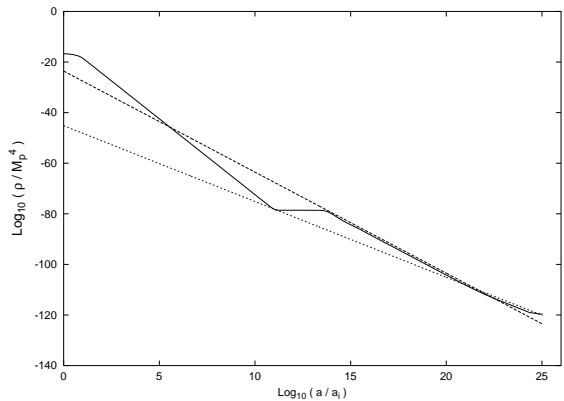}}
\caption{ The post inflationary evolution of energy density for the scalar field (solid line) radiation (dashed line-Born-Infeld matter) and cold dark matter (dotted line) is shown as a function of the scale factor in case of the model described by 
Eq (\ref{cosine}) with $V_0\simeq 1.6 \times 10^{-30} M_p$,
$\tilde{\alpha}=25$ and $p=0.2$. Briefly after the brane effects are over, the field energy density $\rho_{\phi}$ enters the kinetic regime and then overshoots the background. After a short while it turns to the background and tracks it for a long time till the field
reaches the origin and rapid oscillations commence in the system allowing the universe to accelerate at late times. The evolution
is shown from the end of inflation ($a_i \equiv a_{end}=1$) to the present epoch (the value of scale factor $a \simeq 10^{25}$ corresponds to the present epoch)
}
\label{den}
\end{figure}

\begin{figure}
\resizebox{3.0in}{!}{\includegraphics{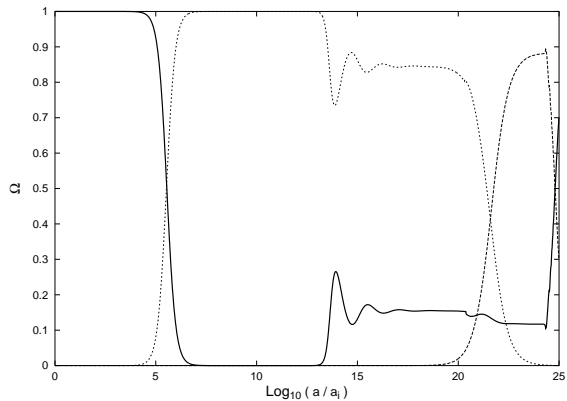}}
\caption{The dimensionless density parameter is plotted as a function of scale factor for the model described in figure \ref{den}.
Solid line corresponds to scalar field, dashed to Born-Infeld radiation and dotted to cold dark matter. Late time oscillations with
negative average equation of state parameter give rise to current epoch of cosmic acceleration with $\Omega_{\phi}=0.7$ and
$\Omega_{m}=0.3$.
}
\label{Omega}
\end{figure}

As discussed above, the scalar field with exponential potential (\ref{exppot}) along with the Born-Infeld matter leads to a viable
evolution at early times. We should, however, ensure that the scalar field becomes quintessence at late times. Indeed, the {\it cosine 
hyperbolic} potential (\ref{cosine}) changes to power law like behavior near the origin $\phi \simeq 0$ giving rise to field oscillations
at late times. For a particular choice of power law the average equation of state may turn negative\cite{turner,sahni}
\begin{equation}
\left< w_{\phi} \right> \simeq \left <{{ {\dot{\phi}^2 \over 2}-V(\phi)} \over {{\dot{\phi}^2 \over 2}+V(\phi)}} \right >={{p-1} \over{p+1}}
\label{stateeq}
\end{equation}
As a result the scalar field energy density and the scale factor have the following behavior
$$\rho_{\phi} \propto  a^{-3(1+\left< w \right>)},~~~~~~~~a \propto   t^{{ 2 \over 3}(1 + \left< w \right>  )^{-1}}. $$
The average equation of state
$ \left<w(\phi) \right> <-1/3$ for $p < 1/2$ allowing the scalar field to play the role of dark energy. We have numerically investigated the model
(with radiation sourced by the Born-Infeld matter) by including the standard cold dark matter. For a particular choice of parameters
$V_0$, $\tilde{\alpha}$ and $p$, the results are displayed in figures \ref{den}, \ref{Omega}. We observe that briefly after the inflation
has ended, the scalar field enters the kinetic regime and scalar field energy density overshoots the background (radiation). Once the
background is overshot, $\rho_{\phi}$ turns towards it and gets locked ($w_{\phi}=-1$) for some time. Tracking commences in the radiative regime 
and continues for a substantially long period. In the model described by Eq (\ref{cosine}), the scalar field is moving from
large values towards the origin. Due to a specific character of the potential, fast oscillations build up in the system as $\phi$ approaches the
origin allowing $\rho_{\phi}$ to scale slower than the background energy density. As a result $\rho_{\phi}$ moves towards the background,
overtakes it and becomes dominant to account for the late time accelerated expansion of the universe. In contrast to the `quintessential
inflation' based upon the gravitational particle production mechanism where the scalar field spends long time in the kinetic regime and makes deep undershoot followed by long locking period with very brief tracking, the scalar field in the present scenario tracks
the background for a very long time (see figure \ref{den}). This pattern of evolution is consistent with the thermal history of the universe. We note that 
`quintessential inflation' can also be implemented by inverse power law potentials.\par
Let us summarize below the chronology of main events in the model: (i) For most of the time during inflation driven by the scalar field, the Born-Infeld 
matter energy density mimics cosmological constant like behavior being sub-dominant to the scalar field energy density $\rho_{\phi}$. shortly before the inflation
comes to end, $\rho_{BI}$ drops below $6\lambda_b$ and begins to scale as radiation. At the end of inflation, the Born-Infeld
radiation density is smaller than $\rho_{\phi}$ by several orders of magnitudes. (ii) After the inflation has ended, the field
energy density scales with the scale factor as $1/a$ and brane effects persists for sometimes. Due to the steepness of the potential,
the kinetic energy of the scalar field fast increases making the potential energy term in the field evolution equation irrelevant
and allowing the commencement of the kinetic regime. (iii) The scalar field continues to be in the kinetic regime for quite
sometime and overshoots the background (radiation). (iv) Once this happens, the dominant contribution to the Friedman equation
comes from the background. This, in turn, allows $\rho_{\phi}$ to turn towards the background leading to locking period. (v) The
scalar field starts rolling down its potential again. The field $\phi$ which is is evolving from larger values towards the origin
is still far away from $\phi \simeq 0$ at this stage. The scalar field then joins the scaling solution which is an attractor of
the system with exponential potential (\ref{exppot}) and continues to track the back ground till $\phi$ approaches the origin. (vi) 
Once  $\phi \simeq 0$, fast oscillations build up leading to $\left<w_{\phi} \right> <-1/3$ and enabling the scalar field to play the role of
dark energy (quintessence) at late times.

\section{Summary and discussion}

In this paper we have discussed a scenario of quintessential inflation based on {\it hybrid brane worlds }. In this model radiation originates from the Born-Infeld action while the scalar field action is supposed to be of the ordinal scalar field. It is shown that the scalar field
in this scenario plays the dual role of inflaton at early epochs and dark energy at late times. The model successfully overcomes
the difficulties associated with the generation of large gravity wave background and is consistent with the nucleo-synthesis constraint. Interestingly the model does not require any additional fine tuning of the parameters.\par
We should, however, comment on the following important issues associated with the scenario: (i) In our model the contribution to energy density after inflation comes
 from inflaton as well as from the Born-Infeld matter. The Born-Infeld
 matter dominates the energy density of the radiation. We have accounted
 for the scalar field fluctuations only and assumed the validity of our
 result similar to the single field inflation models. As the Born-Infeld
 matter dominates the energy density of radiation after inflation, it is pertinent to study 
 also fluctuations in the Born-Infeld matter to check the
 reliability of primordial density perturbations on super-horizon scales
 due to inflaton fluctuations. The fluctuations in Born-Infeld matter may
 or may not be important. We also think that inflation/quintessence may
emerge naturally from D-brane physics, say instability of D-$\bar D$ and so
on. We hope to address these issues in our future work. (ii) The ad hoc inclusion of standard cold dark
matter and (iii) the absence of baryons in the model. One could introduce an additional scalar field with the same potential
as (\ref{cosine}) to mimic the cold dark matter. Indeed, the scalar field behaves like pressure less dust for $p=1$
(see Eq (\ref{stateeq})) and by suitably adjusting the parameters in the potential of the new field, in principal, one could
make the field behave like CDM. Regarding the baryons, the Born-Infeld action, as mentioned above, in general also includes
Fermi fields. The energy density of Fermions scales as radiation in the early universe and could lead to the whole history of our
universe including  matter dominated era. These are important issues which, in our opinion, deserve further investigation.

\section*{Acknowledgments} We are thankful to Yun-Song Piao, S. Panda,Takeo Moroi,
 E. Guendelman and O. Bertolami for useful comments. 
One of us (MS) thanks V. Sahni ,Parampreet Singh, T Padmanabhan and Kandaswamy Subramania for helpful discussions.
The work of TS was supported by Grant-in-Aid for Scientific
Research from Ministry of Education, Science, Sports and Culture of
Japan(No. 13135208, No.14740155 and No.14102004).

{99}


\begin{thebibliography}{99}

\bibitem{vilenkin}P.~J.~Peebles and A.~Vilenkin,
Phys.\ Rev.\ D {\bf 59}, 063505 (1999).
\bibitem{lid}E.J. Copeland, A. R. Liddle and J. E. Lidsey, Phys.Rev. D64 (2001) 023509[astro-ph/0006421].
\bibitem{liddle}
E.J. Copeland, A. R. Liddle and J. E. Lidsey, Phys. Rev. D{\bf 64}, 023509(2001).
\bibitem{lidsey1} 
G. Huey and Lidsey, Phys. Lett. B{\bf 514},217(2001).
\bibitem{tarun} 
V. Sahni, M. Sami and T. Souradeep, Phys. Rev. D{\bf 65},023518(2002).
\bibitem{majumdar} A. S. Majumdar, Phys.Rev. D{\bf 64} (2001) 083503.
\bibitem{lidsey2} 
J. E. Lidsey, T. Matos and  L. Arturo Urena-Lopez, Phys.Rev. D{\bf 66},023514,(2002).
\bibitem{nunes}
N. J. Nunes and E. J. Copeland, Phys. Rev. D{\bf 66},043524(2002). 
\bibitem{ford}
L.H. Ford, Phys. Rev. D{\bf 35}, 2955 (1987).
\bibitem{spokoiny} 
B. Spokoiny, Phys. Lett. B{\bf 315}, 40 (1993).
\bibitem{randall}
L.Randall and R. Sundrum , Phys. Rev. Lett.{\bf 83},4690 (1999);
L. Randall and R. Sundrum, Phys. Rev. Lett. {\bf 83}, 3370 (1999)
\bibitem{shiromizu}
T. Shiromizu, K. Maeda and M. Sasaki, Phys. Rev. D{\bf 62},024012 (2000); 
M. Sasaki, T. Shiromizu and K. Maeda, Phys. Rev. D{\bf 62},024008 (2000).
\bibitem{braneindustry}  
J.M. Cline, C. Grojean and G. Servant, Phys. Rev. Lett. {\bf 83} 4245 (1999);P.~Bin\'etruy, C.~Deffayet and D.~Langlois,
Nucl.\ Phys.\ B {\bf 565} (2000) 269>, [arXiv:hep-th/9905012];
C.~Csaki, M.~Graesser, C.~F.~Kolda and J.~Terning, Phys.\ Lett.\ B {\bf
462}, 34 (1999) [arXiv:hep-ph/9906513]; P.~Bin\'etruy, C.~Deffayet,
U.~Ellwanger and D.~Langlois, Phys.\ Lett.\ B {\bf 477} (2000) 285
[arXiv:hep-th/9910219]; 
A. Chamblin and H. S. Reall, Nucl. Phys. B{\bf 562},133(1999); 
T. Nihei, Phys. Lett. B{\bf 465}, 81(1999); 
N. Kaloper, Phys. Rev. D{\bf 60},123506(1999); 
H. B. Kim and H. D. Kim, Phys. Rev. D{\bf 61}, 064003(2000); 
P. Kraus, JHEP {\bf 9912}, 011(1999);
D. Ida, JHEP {\bf 0009}, 014(2000); 
E. E. Flanagan, S.H.H. Tye and I. Wasserman, Phys. Rev. D{\bf 62}, 044039(2000);
A. Chamblin, A. Karch, A. Nayeri, Phys. Lett. B{\bf 509},163(2001); 
P. Bowcock, C. Charmousis and R. Gre-gory, Class. Quant. Grav. {\bf 17},4745(2000); 
S. Mukohyama, Phys. Lett. B{\bf 473}, 241(2000); 
J. Garriga and M. Sasaki, Phys. Rev. D{\bf 62},043523(2000); 
S. Mukohyama, T. Shiromizu and K. Maeda, Phys. Rev. {\bf D62}, 024028(2000); 
S. Nojiri, S. D. Odintsov and S. Ogushi, hep-th/0205187;
D. Langlois, hep-th/0209261;
M. Sami, Grav.Cosmol. {\bf 7},228 (2001).
\bibitem{gman} E. I. Guendelman, Mod.Phys. Lett. A{\bf 14}, 1043(1999); E. I. Guendelman and O. Katz,
gr-qc/0211095. 
\bibitem{curvaton}
A. R. Liddle and  L. Arturo Urena-Lopez, astro-ph/0302054.
\bibitem{curvaton1} 
D. H. Lyth and D. Wands, Phys. Lett. B{\bf 524},5(2002); Phys. Rev. D{\bf 67},023503(2003); 
D. H. Lyth, C. Ungareli and D. Wands, Phys. Rev. D{\bf 67},023503(2003);  
K. Dimopoulos, D. H. Lyth, A. Notari and A. Riotto, hep-ph/0304050.
\bibitem{curvatonJ} Takeo Moroi and Tomo Takahashi,Phys.Lett.B522, 221(2001); Erratum-ibid.B{\bf539}303(2002);
Takeo Moroi and Tomo Takahashi,  Phys.Rev.D{\bf 66}, 063501(2002);  Motoi Endo, Masahiro Kawasaki and Takeo Moroi, hep-ph/0304126.


 \bibitem{curvaton2}
B. Feng and M. Li, astro-ph/0212213.
\bibitem{curvaton3} 
K. Dimopoulos, astro-ph/0212264.
\bibitem{bi1}  
T. Shiromizu, T. Torii and T. Uesugi, hep-th/0302223.
\bibitem{bi2} 
A. A. Tseytlin, hep-th/9908105.
\bibitem{bi3} 
M. R. Garousi, Nucl. Phys. B{\bf 584},284 (2000)[hep-th/0003122].
\bibitem{bi3panda} E.A. Bergshoeff, M. de Roo, T.C. de Wit, E. Eyras and S. Panda, JHEP 0005 (2000) 009. 
\bibitem{bi4} 
C. Kim, H. B. Kim, Y. Kim and O-K. Kwon, hep-th/0301142; Chanju Kim, Hang Bae Kim, Yoonbai Kim, O-Kab Kwon, JHEP 0303 (2003) 008.
\bibitem{salim}V. A. De Lorenci, R. Klippert, M. Novello and J. M. Salim, Phys. Rev. D{\bf 65},
063501(2002).
\bibitem{tytgat}
A. Mazumdar, S. Panda, A. Perez-Lorenzana,
Nucl. Phys. B 614 (2001) 101;
M. Fairbairn and M.H.G. Tytgat, Phys. Lett. B{\bf 546},1 (2002); 
A. Feinstein, Phys. Rev. D{\bf 66},063511(2002);G Shiu,  and I. Wasserman,
Phys.Lett. B541 (2002) 6.  G Shiu, S.-H. Henry Tye, I. Wasserman, Phys. Rev. D67 (2003) 083517; 
D. Choudhury, D.Ghoshal, D. P. Jatkar and S. Panda, Phys. Lett. B{\bf 544},231(2002);
T. Padmanabhan, Phys.Rev. D{\bf 66},021301 (2002);
L. Kofman and A. Linde, JHEP {\bf 0207},004 (2002);Jian-gang Hao, Xin-zhou Li, hep-th/0209041; Xin-zhou Li, Jian-gang Hao, Dao-jun Liu,hep-th/0204252; 
M. Sami, P. Chingangbam and T. Qureshi, Phys. Rev. D{\bf 66},043530 (2002); 
M. Sami, Mod. Phys. Lett. A{\bf 18},691 (2003); 
M. Sami, P. Chingangbam and T. Qureshi, hep-th/0301140; 
S. Mukohyama, Phys. Rev. D{\bf 66},024009(2002); 
G. Felder, L. Kofman and A. Starobinsky, JHEP {\bf 0209}, 026,(2002);M. C. Bento, O. Bertolami and A.A. Sen, hep-th/020812;  M.C. Bento, O. Bertolami., A.A. Sen, Phys.Rev.D67:023504,2003;
 gr-qc/0204046; M.C. Bento, O. Bertolami., Phys.Rev.D65:063513,2002; astro-ph/0111273; B. Wang, E. Abdalla and R. -K. Su, hep-th/0208023;
Yun-Song Piao, Rong-Gen Cai, Xinmin Zhang, Yuan-Zhong Zhang, Phys.Rev. D{\bf 66} (2002)121301;
Yun-Song Piao, Qing-Guo Huang, Xinmin Zhang, Yuan-Zhong Zhang,hep-ph/0212219
;Xin-zhou Li, Dao-jun Liu, Jian-gang Hao; hep-th/0207146;Zong-Kuan Guo, Yun-Song Piao, Rong-Gen Cai, Yuan-Zhong Zhang,  hep-ph/0304236;  T. Padmanabhan, hep-th/0212290;
J. M. Cline and H. Firouzjahi, hep-th/0301101;  Tomohiro Matsuda, hep-ph/0302035
; Tomohiro Matsuda, hep-ph/0302078; J.S.Bagla, H.K.Jassal, T.Padmanabhan, Phys.Rev. D{\bf 67} (2003) 063504.
 T. Padmanabhan, T. Roy Choudhury,  Phys.Rev. D{\bf 66} (2002) 081301; Ph.Brax, J.Mourad, D.A.Steer, hep-th/0304197; A. Das, A. DeBenedictis, gr-qc/0304017;
; Chanju Kim, Yoonbai Kim, Chong Oh Lee,
hep-th/0304180; Mahbub Majumdar, Anne-Christine Davis, hep-th/0304226; D. Choudhury, D. Ghoshal,  Dileep P. Jatkar (1), S. Panda, hep-th/0305104;Jian-gang Hao, Xin-zhou Li,hep-th/0306033;  Jian-gang Hao, Xin-zhou Li, hep-th/0305207; Shin'ichi Nojiri and  Sergei D. Odintsov, hep-th/0306212;  D. Bazeia, F.A. Brito and J.R.S. Nascimento, hep-th/0306284. 



 \bibitem{g1} 
L. P. Grishchuk, Sov. Phys. JETP {\bf 40}, 409 (1975). 
A pre-inflationary analysis of relic gravity waves created during a 
stiff-matter phase can be found in: L. P. Grishchuk, Ann. (N.Y.) Acad. Sci. {\bf 302}, 439 (1977); 
B. L. Hu and L. Parker, Phys. Lett. {\bf A63}, 217 (1977).
\bibitem{g2} 
A. A. Starobinsky, JETP Lett., {\bf 30}, 682 (1979).
\bibitem{g3}
B. Allen, Phys. Rev. {\bf D37}, 2078 (1988).
\bibitem{g4} 
V. Sahni, Phys. Rev. D{\bf 42}, 453 (1990).
\bibitem{g5}
D. Langlois, R. Maartens and D. Wands, Phys. Lett. {\bf B489},259(2000).
\bibitem{g6} 
The generation of `blue' gravity wave spectra in inflationary models with a stiff post-inflationary equation of state has also
been discussed in: M. Giovannini, Phys. Rev. {\bf D58}, 083504 (1998). The production of relic gravity waves in quintessential
inflationary models is treated in detail in : M. Giovannini, Phys. Rev. {\bf D60}, 123511 (1999); Class. Quant. Grav. {\bf 16}, 2905
(1999).
\bibitem{g7} 
A. Buonanno, TASI Lectures on Gravitational Waves from the Early Universe, gr-qc/0303085. 
\bibitem{g8} Tsutomu Kobayashi, Hideaki Kudoh and Takahiro Tanaka, gr-qc/0305006. 
\bibitem{sahni} 
V. Sahni and L. Wang, Phys. Rev. {\bf D62}, 103517 (2000).
\bibitem{turner}
M. S. Turner, Phys. Rev. {\bf D28} ,1243 (1983); 
T. Damour and V. F. Mukhanov , Phys. Rev. Lett. {\bf 80}, 3440 (1998);
A. R. Liddle and A. Mazumdar, Phys. Rev. D{\bf 58},083508(1998); 
V. Cardenas and G. Palma ,Phys. Rev. D{\bf 61} ,027301(1999); 
J. Lee et al, Phys. Rev. D{\bf 61} ,027302 (1999);
M. Sami, Grav. Cosmol. {\bf 8}, 309(2003); 
M. Sami and T. Padmanabhan, Phys. Rev. D{\bf 67}, 083509(2003)[hep-th/0212317].

\end{thebibliography}
\end{document}